\documentstyle[draft,epsf]{mn}

\def\gsim{\mathrel{\raise0.35ex\hbox{$\scriptstyle >$}\kern-0.6em 
\lower0.40ex\hbox{{$\scriptstyle \sim$}}}}
\def\lsim{\mathrel{\raise0.35ex\hbox{$\scriptstyle <$}\kern-0.6em 
\lower0.40ex\hbox{{$\scriptstyle \sim$}}}}

\def\msun{{\rm M$_{\odot}$}}

\title{
Down-Sizing in Galaxy Formation at $z\sim1$ in the
Subaru/XMM-Newton Deep Survey (SXDS)
}

\author[Kodama et al.]{
\parbox[t]{\textwidth}{
Tadayuki Kodama$^{1}$, Toru Yamada$^{1}$, 
Masayuki Akiyama$^2$, Kentaro Aoki$^2$,\\
Mamoru Doi$^4$, Hisanori Furusawa$^2$,
Tetsuharu Fuse$^2$, Masatoshi Imanishi$^1$,\\
Cathy Ishida$^2$, Masanori Iye$^2$,
Masaru Kajisawa$^2$, Hiroshi Karoji$^2$,\\
Naoto Kobayashi$^4$, Yutaka Komiyama$^2$,
George Kosugi$^2$, Yoshitomo Maeda$^5$,\\
Satoshi Miyazaki$^2$, Yoshihiko Mizumoto$^1$,
Tomoki Morokuma$^4$,\\
Fumiaki Nakata$^6$, Junichi Noumaru$^2$,
Ryusuke Ogasawara$^2$, Masami Ouchi$^3$,\\
Toshiyuki Sasaki$^2$, Kazuhiro Sekiguchi$^2$,
Kazuhiro Shimasaku$^3$,\\
Chris Simpson$^6$, Tadafumi Takata$^2$,
Ichi Tanaka$^7$, Yoshihiro Ueda$^5$,\\
Naoki Yasuda$^8$, Michitoshi Yoshida$^9$ (The SXDS Team)}
\vspace*{6pt}\\
$^{1}$National Astronomical Observatory of Japan, Mitaka, Tokyo
181--8588, Japan \\
$^{2}$Subaru Telescope, National Astronomical Observatory of Japan,
Hilo, HI 96720, U.S.A \\
$^{3}$Department of Astronomy, School of Science, University of Tokyo,
Bunkyo-ku, Tokyo 113--0033, Japan \\
$^{4}$Institute of Astronomy, University of Tokyo, Mitaka, Tokyo
181--1500, Japan \\
$^{5}$Institute of Space and Astronautical Science,
Sagamihara, Kanagawa 229--8510, Japan \\
$^{6}$Department of Physics, University of Durham, South Road, Durham
DH1 3LE, UK \\
$^{7}$Astronomical Institute, Tohoku University, Aoba-ku,
Sendai 980--8578, Japan\\
$^{8}$Institute for Cosmic Ray Research, University of Tokyo, Kashiwa,
Chiba 277--8582, Japan \\
$^{9}$Okayama Astrophysical Observatory, National Astronomical Observatory of
Japan, Asakuchi-gun, Okayama 719--0232, Japan
}

\begin{document}

\label{firstpage}

\maketitle

\date{Accepted for publication in MNRAS Main Journal}

\begin{abstract}

We use the deep wide-field optical imaging data of the Subaru/XMM-Newton
Deep Survey (SXDS) to discuss the luminosity (mass) dependent galaxy
colours down to $z'$=25.0 (5$\times$10$^9$$h_{70}^{-2}$M$_{\odot}$) for
$z\sim1$ galaxies in colour-selected high density regions. We find an
apparent absence of galaxies on the red colour--magnitude sequence below
$z'\sim24.2$, corresponding to $\sim$$M^*$+2 ($\sim$10$^{10}$M$_{\odot}$)
with respect to passively evolving galaxies at $z\sim1$.  Galaxies brighter
than $M^*$$-$0.5 (8$\times$10$^{10}$M$_{\odot}$), however, are
predominantly red passively evolving systems, with few blue star forming
galaxies at these magnitudes.

This apparent age gradient, where massive galaxies are dominated by old
stellar populations while less massive galaxies have more extended star
formation histories, supports the `down-sizing' idea where the mass of
galaxies hosting star formation decreases as the Universe ages.  Combined
with the lack of evolution in the shape of the stellar mass function for
massive galaxies since at least $z\sim1$, it appears that galaxy formation
processes (both star formation and mass assembly) should have occurred in
an accelerated way in massive systems in high density regions, while these
processes should have been slower in smaller systems.  This result provides
an interesting challenge for modern CDM-based galaxy formation theories
which predict later formation epochs of massive systems, commonly referred
to as ``bottom-up''.
\end{abstract}


\section{Introduction}

Galaxy properties depend strongly on the mass of the system.  Based on the
122,808 galaxies drawn from the {\it Sloan Digital Sky Survey (SDSS)},
Kauffmann et al. (2003) have recently shown an interesting bimodality of
local galaxy properties separated at a stellar mass of
$\sim$3$\times$10$^{10}$\msun.  In particular, in contrast to massive
galaxies which are dominated by old stellar populations showing little sign
of recent star formation, less massive galaxies have a much larger
contribution from young stars and a significant fraction of these low mass
galaxies are likely to have experienced recent starbursts (see also
Gavazzi \& Scodeggio 1996; Baldry et al.\ 2004).
The morphological signatures also
depend on the mass or luminosity, with massive galaxies showing
centrally-concentrated light profiles (early-type/bulge morphologies), and
less massive galaxies showing less concentrated profiles (late-type/disk
morphologies; e.g., Kauffmann et al.\ 2003; Treu et al.\ 2003).

Based on the $z\gsim1$ galaxies in the {\it Hawaii Deep Fields}, Cowie et
al.\ (1996) have suggested that the most massive galaxies form earliest in
the Universe, and star formation activity is progressively shifted to
smaller systems, although their data are limited to galaxies brighter than
$M^*$.  They have termed this phenomenon `down-sizing' in star forming
galaxies. This apparent age gradient as a function of mass, where the
massive galaxies are uniformly old while the less massive galaxies tend to
be younger, appears to be at odds with cold dark matter (CDM) models of the
Universe. In these models, galaxies form in a `bottom-up' or hierarchical
manner, with small systems collapsing first and massive galaxies forming
later via the assembly of these small systems.

Motivated by this interesting and fundamental puzzle in galaxy formation,
we have chosen to investigate the mass dependence of galaxy properties at
high redshift ($z\sim1$) in more detail by studying much lower mass systems
than previous work. To do this, we have used a statistically large sample
of galaxies drawn from the Subaru/XMM-Newton Deep Survey (Sekiguchi et al.\
2004), which has the unique advantage of depth ($z'$=25.2 at 5--8$\sigma$)
and width (1.2 deg$^2$) achieved via the wide-field (30') camera
Suprime-Cam on the 8.2-m Subaru Telescope.

Bell et al.\ (2004) have recently analysed the redshift-dependent colour
distributions of $\sim$25000 galaxies over 0.78 deg$^2$ based on the
COMBO-17 optical survey (Wolf et al.\ 2003). This survey, however, is
$\sim$3 magnitudes shallower than the SXDS, reaching only to $\sim$$M^*$ at
$z\sim1$. Our data therefore provides the first opportunity to study the
properties of faint galaxies at high redshift.

We adopt the cosmological parameters ($H_0$, $\Omega_m$,
$\Omega_{\Lambda}$)=(70, 0.3, 0.7) throughout this paper, and define
$h_{70}$ as $H_0$/(70~km s$^{-1}$Mpc$^{-1}$).  With this parameter set, 1
arcmin corresponds to 0.48~Mpc at $z\sim1$.  All the magnitudes in this
paper will be given in the AB-magnitude system.

We structure the paper as follows.  After an introduction in \S1, we
briefly describe in \S2 the Subaru imaging data upon which the following
analyses are based. In \S3, we identify the high density regions at
$z\sim1$ by using the red sequence colour slice technique and determine the
galaxy demographics in these regions by subtracting off foreground and
background contaminations in a statistical manner. We investigate the
photometric properties of this statistical sample of $z\sim1$ galaxies in
\S4, with particular emphasis on the faint end. A discussion of our results
and conclusions are given in \S5 and \S6, respectively.

\section{Observation and Data Reduction}

Detailed information on the observations, the data set and the data
reduction of the Subaru/XMM-Newton Deep Survey (SXDS) project 
will be fully described in Sekiguchi et al.\ (2004)
and Furusawa et al.\ (2004).

Here we only briefly summarise the data on which our analyses are based.
The SXDS Field is located at RA=02:18:00, Dec=$-$05:00:00 (J2000). This is
a multi-wavelength survey programme being conducted using a range of
facilities including {\it Subaru/Suprime-Cam/FOCAS}, {\it UKIRT/WFCAM}, {\it
XMM-Newton}, {\it VLA}, and {\it JCMT/SCUBA}.

As a major part of this survey, we have taken deep Subaru optical images
of a 1.2 deg$^2$ field using 5 pointings of the wide-field camera,
Suprime-Cam (34$'$$\times$27$'$), which covers most of the coordinated {\it
XMM-Newton} deep survey fields.
We have $B$, $R$, $i'$ and $z'$-band images for each pointing and the depths
are slightly different among the pointings, but go uniformly down to
$B\sim27.4$, $R\sim26.8$, $i'\sim26.4$ and $z'\sim25.2$ 
in a 2-arcsec aperture at $\ge$5$\sigma$.
The typical seeing sizes in the combined frames are 0.78--0.9$''$.

We have used SExtractor Version 2.2.2 (Bertin \& Arnouts 1996) to detect
objects in the $z'$-band and perform photometry on the sources in all the
bands.  We use MAGBEST as an estimate of the total $z'$ magnitude, and the
1.8$''$ diameter aperture magnitudes to derive colours such as $R-z'$ and
$i'-z'$.  We exclude stellar objects from our analyses by rejecting sources
for which the CLASS\_STAR index is larger than 0.8 in the $z'$-band image.

Due to concerns that our rejection of compact objects may exclude some low
luminosity galaxies at high redshift, which could affect our later
discussion of the deficit of faint red galaxies, we have repeated our
analysis without performing this star--galaxy separation.  We find that
there is little effect on the derived colour--magnitude diagram, luminosity
function and stellar mass function of $z\sim1$ galaxies, except for the
retention of some obvious stars which are not properly removed in the
statistical subtraction process (\S3.2). These objects are clearly too blue
($R-z'<0.6$) and too bright ($z'<21.5$) to be real galaxies at $z\sim1$.
The analyses which follow in this paper are therefore based on the sample
of objects with CLASS\_STAR$<$0.8.

\section{Data Analysis}

\subsection{Identification of high density regions at $z\sim1$}

In this paper we focus on galaxies at $z\sim1$, since this is the highest
redshift which can be selected and studied in the rest-frame optical
spectral regime based solely on our optical imaging observations. To
construct a sample of $z\sim1$ galaxies from our data, we first identify
overdense regions at $z\sim1$ using colour selection criteria and then make
a statistical correction for foreground and background contamination using
the galaxies from control fields.  We describe this technique more in
detail below.

We know {\it a priori} that clusters of galaxies contain large excesses of
red early-type galaxies compared to the averaged field, and that these
galaxies show a tight colour--magnitude relationship both locally and at
high redshifts (eg., Visvanathan \& Sandage 1977; Dressler 1980; Butcher \&
Oemler 1984; Bower, Lucey \& Ellis 1992; Ellis et al.\ 1997; Stanford et
al.\ 1998; Kodama et al.\ 1998; van Dokkum et al.\ 1998). We can therefore
search for high density regions at $z\sim1$ by mapping the galaxies within
a narrow colour range around the expected location of the colour--magnitude
sequence at that redshift.  The high density regions are identified as
galaxy clumps on the projected sky after these colour cuts.  This technique
has been applied successfully by many authors (e.g., Gladders \& Yee 2000;
Kodama et al.\ 2001) and recognised as an efficient way of searching for
high density systems in the distant Universe.

\begin{figure}
\begin{center}
  \leavevmode
  \epsfxsize 1.0\hsize
  \epsffile{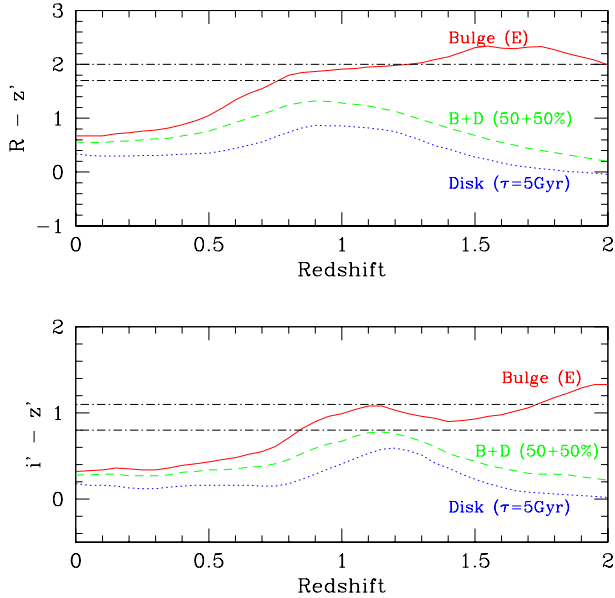}
\end{center}
\caption{
The colour cuts used for selecting the red $z\sim1$ galaxies are indicated
by the horizontal dot-dashed lines delimiting the ranges $1.7<R-z'<2.0$ and
$0.8<i'-z'<1.1$.  Three model colour tracks are plotted as a function of
redshift from Kodama, Bell \& Bower (1999).  The red curves show the colour
tracks for a passively evolving galaxy formed in an instantaneous burst at
$z_f$=5 (Kodama et al.\ 1998).  The blue curves indicate the tracks for
exponentially decaying star formation with an e-folding time of 5~Gyr and
the green curves show the combination of these two models with equal weight
in the rest-frame $B$-band luminosity.  }
\label{fig:colourcut}
\end{figure}

Here we use both $R-z'$ and $i'-z'$ colours. As shown in
Fig.~\ref{fig:colourcut}, the red passive galaxies at $z\sim1$ can be
isolated by using the colour slices defined by $1.7<R-z'<2.0$ and
$0.8<i'-z'<1.1$.  We note that these criteria are expected to produce a
sample of galaxies with a range of redshifts $0.9\lsim z\lsim 1.1$ which
corresponds to the range in distance modulus $\Delta$dm=0.5~mag
(0.2~dex). Including the small k- and e-corrections, the $z'$-band
magnitude can be different by 0.7~mag (0.28~dex) at most. Since we discuss
the photometric properties for our colour-selected sample as a whole, it
should be noted that the scatter in the red sequence and luminosity and
stellar mass functions shown in \S4 are expected to be increased by these
amounts due to the inclusion of galaxies over a range of redshifts.

\begin{table*}
\caption{The $z\sim1$ high density regions identified in the SXDS field.
The relative coordinates are given with respect to the field centre
(RA=02:18:00, Dec=$-$05:00:00).  }
\label{tab:sample}
\begin{center}
\begin{tabular}{cccrrl}
\hline\hline
ID & RA (J2000)  & Dec (J2000) &      dRA   &     dDec   & Note\\
\hline
c1 & 02:18:08.03 & $-$05:01:00 &     2.0$'$ &  $-$1.0$'$ & \\
c2 & 02:19:24.34 & $-$04:51:00 &    21.0$'$ &     9.0$'$ & \\
c3 & 02:20:00.48 & $-$05:09:00 &    30.0$'$ &  $-$9.0$'$ & included due to narrow C-M sequence\\
c4 & 02:20:26.58 & $-$04:48:30 &    36.5$'$ &    11.5$'$ & \\
c5 & 02:17:35.90 & $-$04:32:00 &  $-$6.0$'$ &    28.0$'$ & \\
c6 & 02:16:35.70 & $-$04:49:30 & $-$21.0$'$ &    10.5$'$ & excluded due to two bright stars nearby\\
\hline
\end{tabular}
\end{center}
\end{table*}

\begin{figure}
\begin{center}
  \leavevmode
  \epsfxsize 1.0\hsize
  \epsffile{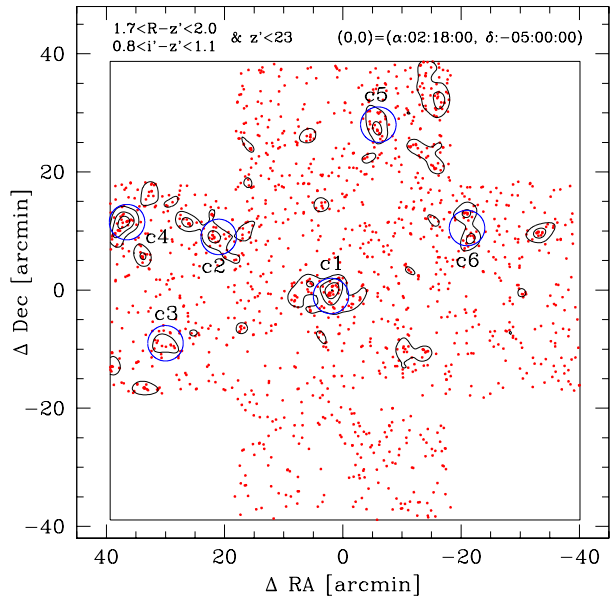}
\end{center}
\caption{
The 2-D distribution of the 1503 `red' galaxies ($z\sim1$) in the SXDS
field (1.2 deg$^2$) selected by the colour and magnitude cuts shown at the
top of the panel.  The contours show the local surface density of the red
galaxies measured by smoothing the galaxy distribution with a 1.2-arcmin
radius Gaussian.  The contours show excesses of 2, 3, 4, and 5$\sigma$
above the average density.  The large open circles (with a radius of 3
arcmin) indicate the $z\sim1$ high density regions identified from this
map.  }
\label{fig:map_z1}
\end{figure}

\begin{figure*}
\begin{center}
  \leavevmode
  \epsfxsize 0.32\hsize
  \epsffile{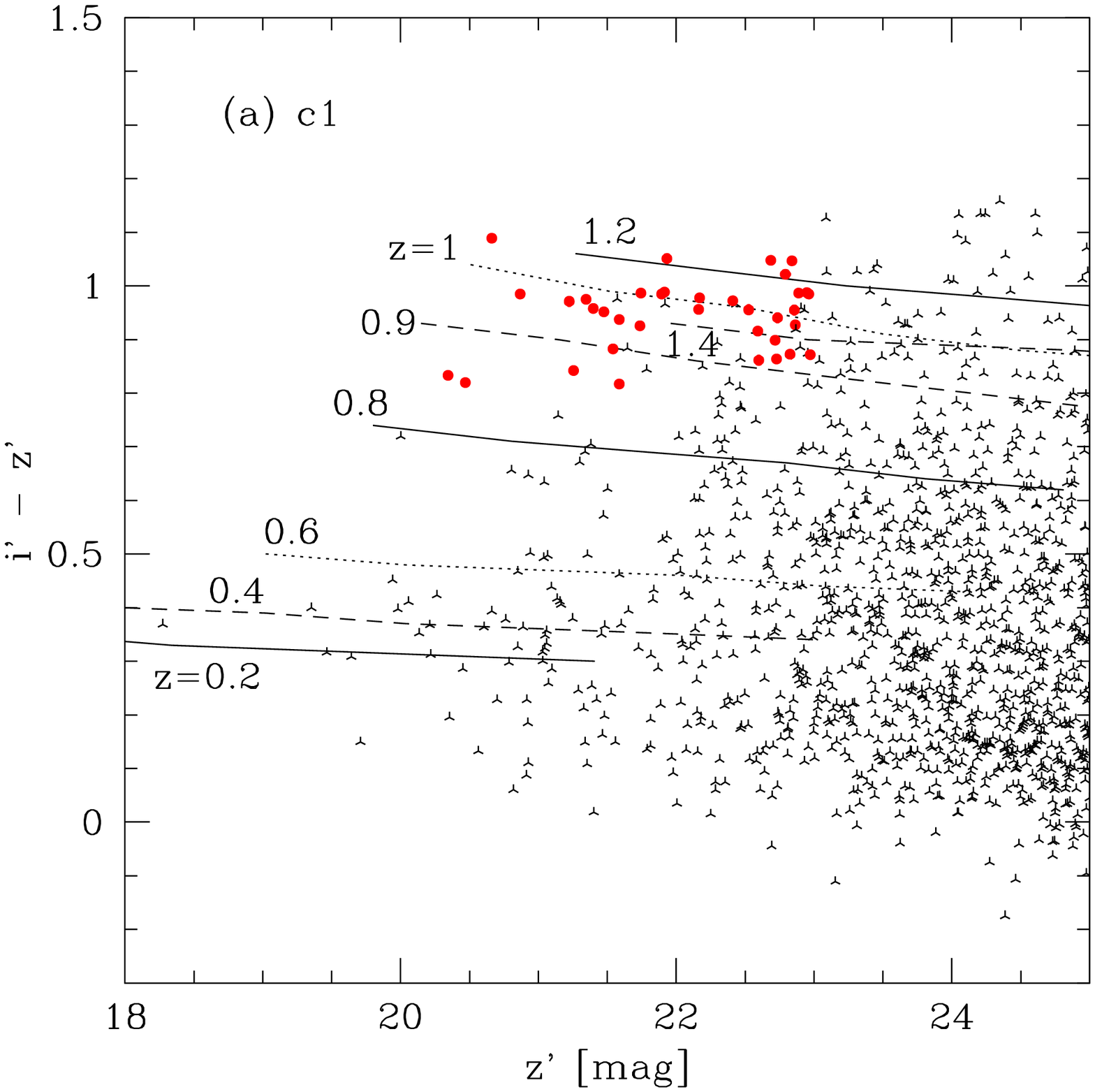}
  \epsfxsize 0.32\hsize
  \epsffile{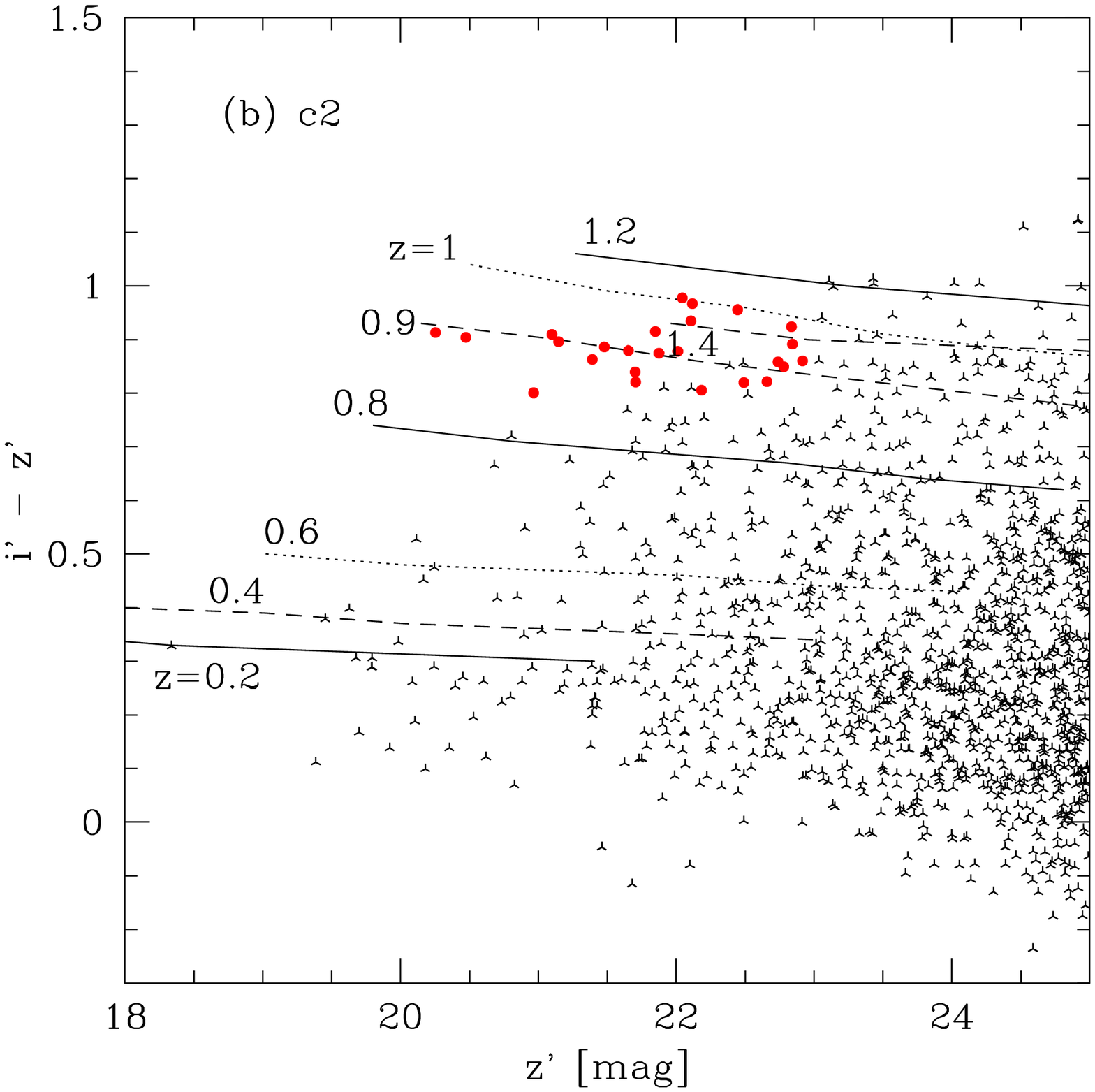}
  \epsfxsize 0.32\hsize
  \epsffile{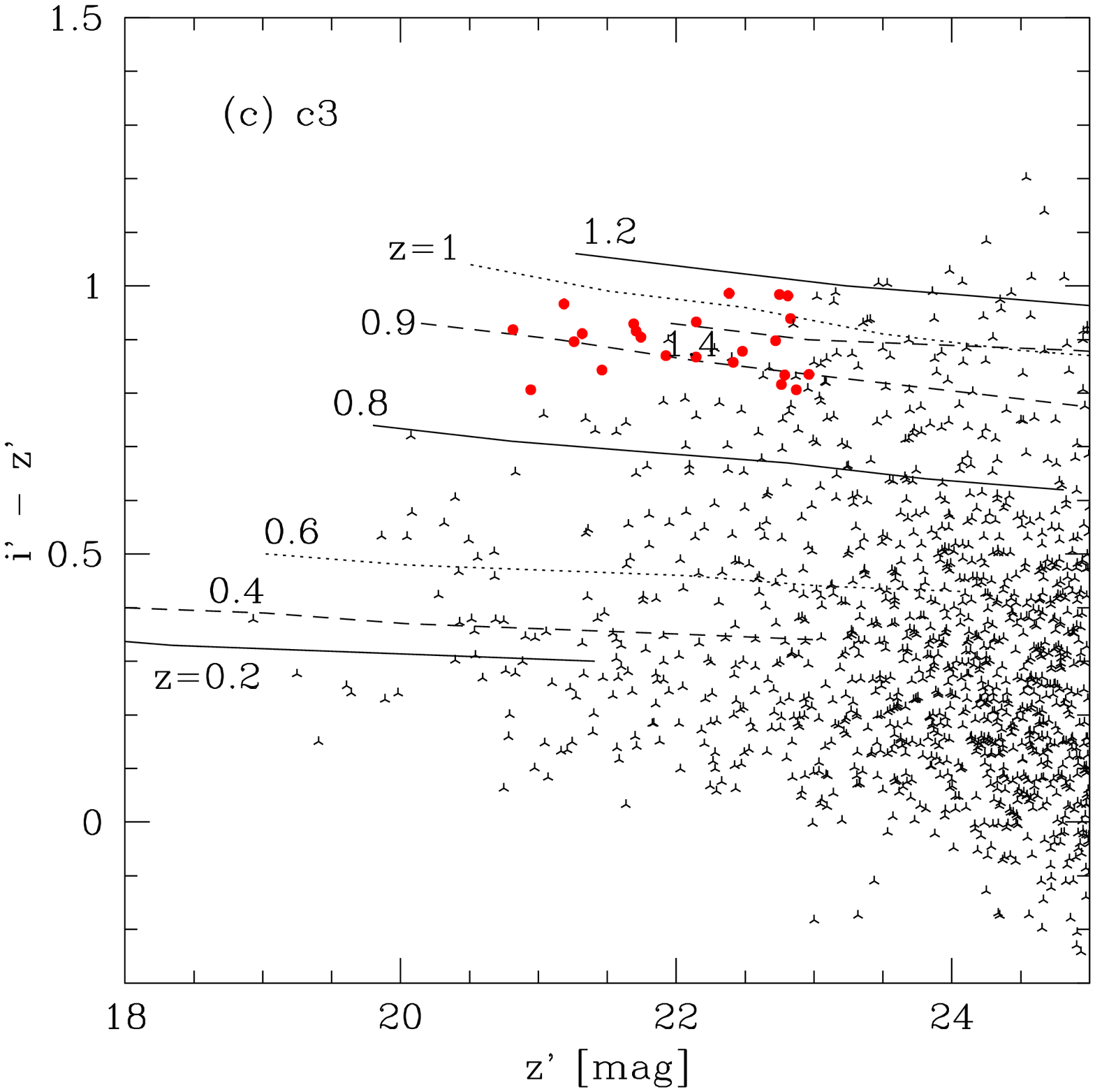}\\
  \vspace{0.2cm}
  \leavevmode
  \epsfxsize 0.32\hsize
  \epsffile{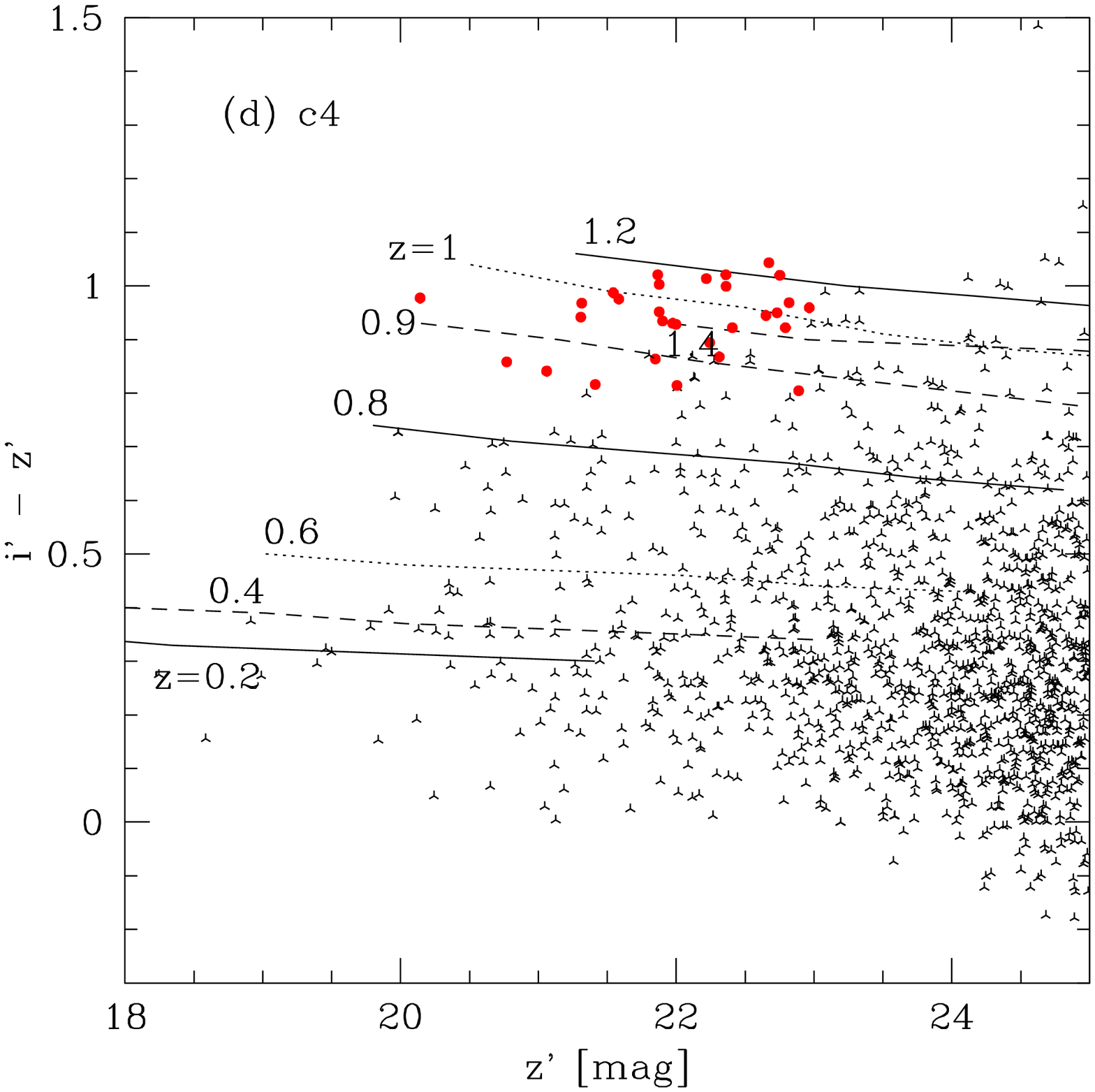}
  \hspace{1.0cm}
  \epsfxsize 0.32\hsize
  \epsffile{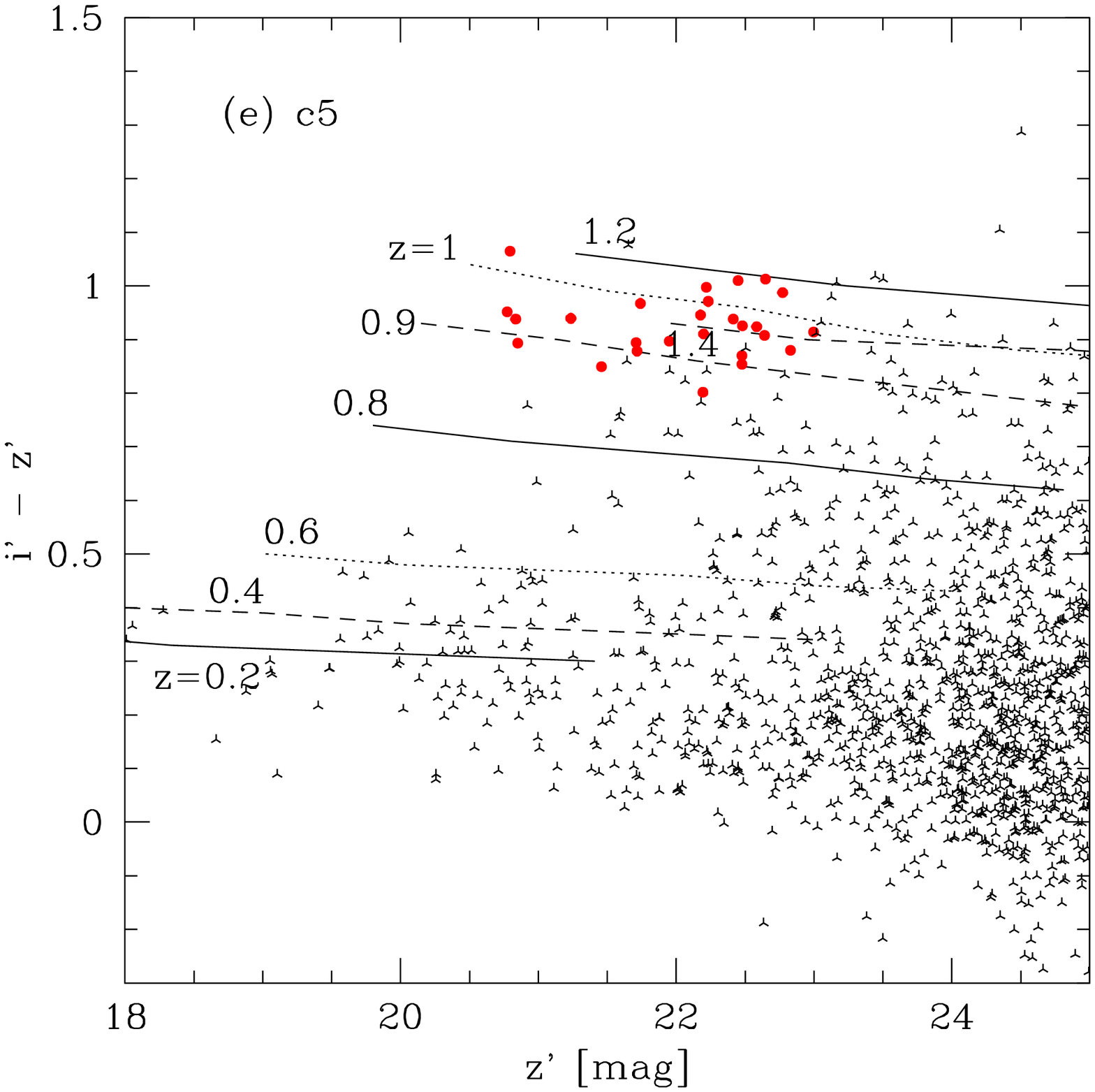}
\end{center}
\caption{Colour--magnitude diagrams for the galaxies within the 3 arcmin
circles of our $z\sim1$ high density regions (c1--c5).  Galaxies with
$z'<23$ and $1.7<R-z'<2.0$ and $0.8<i'-z'<1.1$, on which the overdensities
of the $z\sim1$ red galaxies are based, are shown as filled circles.  }
\label{fig:example}
\end{figure*}

Figure \ref{fig:map_z1} shows the 2-D distribution of the red galaxies
after the colour cuts were applied.  Here we have also limited the galaxies
to those brighter than $z'=23$ to further contrast the red sequence
galaxies against the foreground/background galaxies.  We have selected 6
high density regions (c1--c6), each of which has a 3 arcmin radius as
indicated by the large circles, and we list these in Table
\ref{tab:sample}.  These regions all have a $>$$3\sigma$ excess in surface
density, except for the southeast clump (c3) which is detected at only the
$2\sigma$ level.  Nevertheless, we include this clump since it shows a
tight colour--magnitude sequence and is likely to be a physically
associated system, although the spatial distribution is a bit sparse.  On
the other hand, we do not include the c6 region in our further analysis
since there are two very bright stars nearby ($<$1--2$'$) which are likely
to significantly affect the photometry. We show the colour--magnitude
diagrams for the five selected regions (c1--c5) in Fig.~\ref{fig:example}.

Some spectroscopic observations in the SXDS field were made with FOCAS on
Subaru in October and December 2003.  We obtained spectra for 64
randomly-chosen galaxies from our colour-selected $z\sim1$ sample with
$z'<22$.  Redshifts were obtained for 59 of these, of which 56 (88\% of the
entire spectroscopic sample of 64 galaxies) fall within the redshift
interval of $0.87\le z\le 1.12$.  Our colour criteria for selecting
$z\sim1$ red galaxies are therefore shown to be reliable and allow only
minor contamination. These results do not allow us to estimate the
completeness of our technique, although this is not a concern since we do
not discuss, for example, the absolute number density of $z\sim1$ galaxies
in this paper.

\subsection{Statistical field subtraction}

Since the spectroscopic data for our $z\sim1$ galaxies are still very
sparse (only $\sim$60 galaxies), and that we can reach only down to $z'=23$
in 1--2-hour exposures on Subaru, it is essential to apply a conventional
``statistical'' field subtraction method to investigate the statistical
properties of $z\sim1$ galaxies, especially for faint ($z'>23$)
galaxies. To achieve this, we have selected 15 control fields from the SXDS
image, each with a 3$'$ radius. These fields have been chosen to carefully
avoid bright stars, the 6 $z\sim1$ high density regions (c1--c6), and the
16 foreground cluster candidates which are detected in the XMM-Newton
observations as extended X-ray sources (Fig.~\ref{fig:map_field}).  By
combining all these 15 independent fields, we have a good representative
sample of the average field population over $\sim$424 arcmin$^2$.

\begin{figure}
\begin{center}
  \leavevmode
  \epsfxsize 1.0\hsize
\end{center}
\caption{
The 15 control fields are shown by solid open circles
(with a radius of 3 arcmin).
These regions are selected to avoid the regions severely affected by
bright stars (shown) and the 16 regions of foreground cluster candidates
($0<z<0.8$) where extended X-ray emissions are detected (dotted open circles).
}
\label{fig:map_field}
\end{figure}

We can then subtract this field sample from the target fields (c1--c5) to
isolate the $z\sim1$ galaxies. This works because both the control and
target fields have the same number density of foreground and background
galaxies (statistically speaking), since we have not used any information
about the foreground/background to select the target fields.  To gain better
statistics in the field subtraction process, we have combined the five
$z\sim1$ high density regions and discuss the averaged properties
throughout the rest of this paper.

The actual field subtraction process depends on what we plot.  For example,
to construct the luminosity function in our $z\sim1$ high density regions,
we just straightforwardly subtract the magnitude distribution of the
galaxies in the control fields from that of the high density regions.  In
order to plot a colour--magnitude diagram, however, we must do the
subtraction in two dimensions using colours and magnitudes at the same
time, for which we use the Monte-Carlo technique described in Appendix~A in
Kodama \& Bower (2001). In short, we make grids on the colour--magnitude
diagram (in this paper we use bin sizes of $\Delta$($R-z'$)=0.15 and
$\Delta$$z'$=0.4), and count the number of galaxies in each bin in both the
control and target fields. We scale the former number to match the smaller
area of the target fields, and for each bin we assign a probability that
a galaxy in the target fields should be subtracted as a field galaxy.
We then generate a random number between 0 and 1 for each galaxy and decide
whether it should be removed depending on whether this random number is larger
or smaller than the probability.
If the probability of being a field galaxy exceeds unity
(this happens if the number of galaxies to be subtracted
exceeds the available number of galaxies in a certain bin),
we re-distribute the excess probability to the neighbouring bins.
We apply the same process when producing a $BRz'$ colour--colour diagram,
where we use bin sizes of $\Delta$($R-z'$)=0.15 and $\Delta$($B-z'$)=0.37.

The largest potential problem in the statistical field subtraction comes
from so-called cosmic variance.  Even though we have averaged over the
large areas of the control fields and hence have a good statistical sample
for the averaged field population, the field-to-field variation in such
things as the number density and colour--magnitude distribution still
remains in the limited target field to be processed (the combined $z\sim1$
high density regions comprise a total area of 141 arcmin$^2$).  To assess
the impact of these field variation on the results presented in this paper,
we have divided the control fields into three independent areas, each of
which is a combination of five 3$'$ circles giving a 141 arcmin$^2$ field
of view, the same as our combined target fields. We then use these three
control samples to perform the statistical field subtraction and measured
the scatter in the results. We will discuss the outcome of this process
later in \S4.3 and Fig.~\ref{fig:mf_z1_var}, but the effect is found to be
small since our target field has relatively a wide area by combining the
five regions.

\section{Results}

\subsection{The colour--magnitude and colour--colour diagrams}

\begin{figure*}
\begin{center}
  \leavevmode
  \epsfxsize 0.7\hsize
  \epsffile{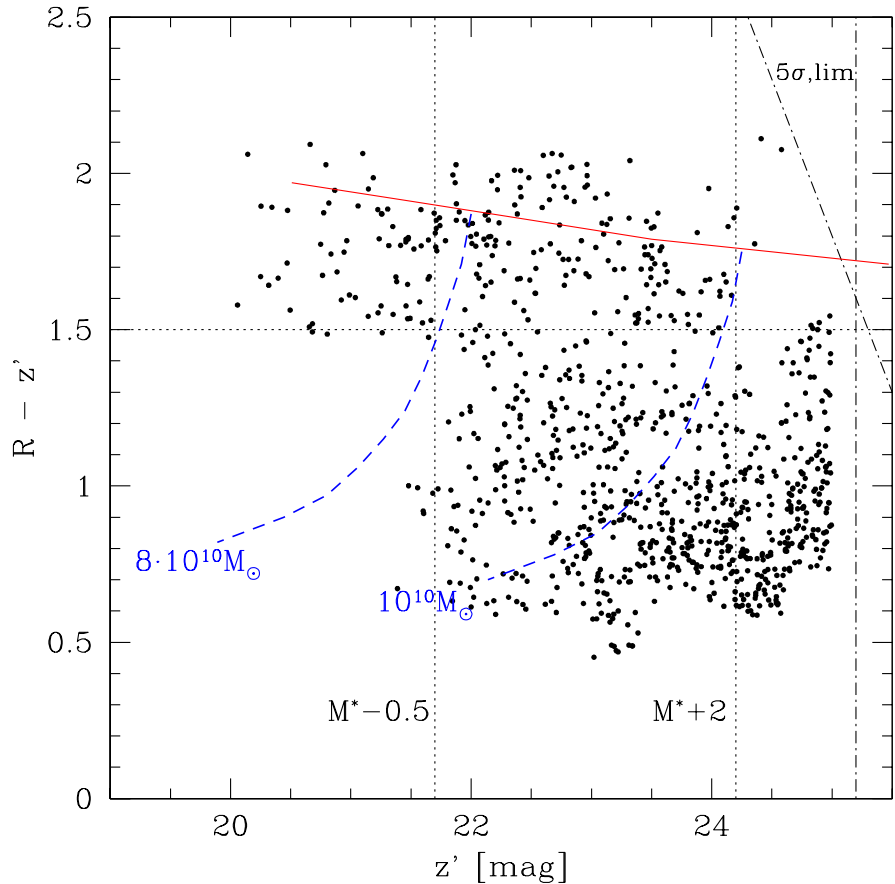}
\end{center}
\caption{
Field-corrected colour--magnitude diagram for the $z\sim1$ galaxies taken
from the combined high density regions (c1--c5).  This is an example of the
Monte-Carlo run and around 900 galaxies are plotted.  The solid line shows
the expected location of the colour--magnitude sequence at $z=1$ assuming
passive evolution since the formation redshift $z_f$=5 (Kodama et al.\
1998).  The dashed curves show lines of constant stellar mass (see \S4.3
for details).  A deficit of red faint galaxies below $z'=24.2$ ($M^*$+2,
right vertical dotted line) or $M_{\rm stars}$=10$^{10}$M$_{\odot}$ is
clearly seen.  Note that this critical magnitude and mass is well above the
5$\sigma$ detection limit shown by the dot-dashed lines.  }
\label{fig:z1_cm}
\end{figure*}

In Fig.~\ref{fig:z1_cm}, we show the field-corrected colour--magnitude
diagram for the $z\sim1$ galaxies in the combined high density regions
(c1--c5). This is a typical Monte-Carlo run for the statistical field
subtraction, and $\sim$900 galaxies are left over after the subtraction
which are expected to lie in the redshift interval of $0.9\lsim z\lsim 1.1$.

\begin{figure}
\begin{center}
  \leavevmode
  \epsfxsize 1.0\hsize
  \epsffile{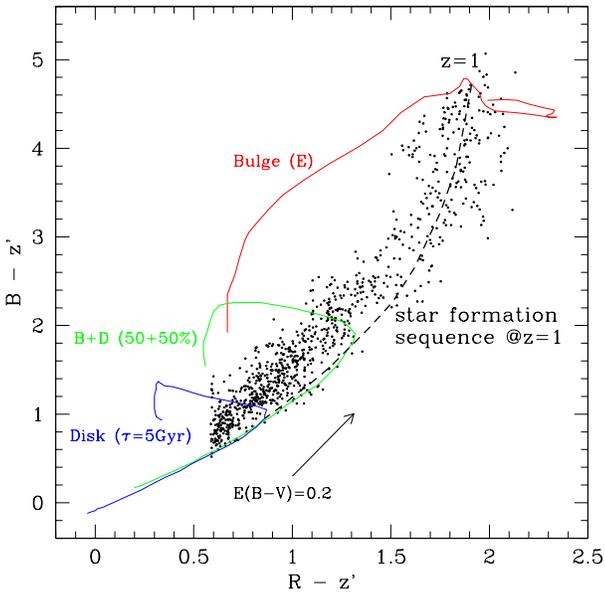}
\end{center}
\caption{
A colour--colour diagram for the same field-corrected $z\sim1$ galaxies as
were plotted in Fig.~\ref{fig:z1_cm}.  The three models plotted here are
the same as those in Fig.~\ref{fig:colourcut}, with redshift increasing
from $z$=0 to 2 in a clockwise direction along the curves.  The internal
reddening vector for $z\sim1$ star forming galaxies (Calzetti et al.\ 2000)
is shown by an arrow.  Most of the galaxies are located along the star
forming locus at $z\sim1$ as shown in the dashed curve, where star
formation rate is changing along the curve, although a slight offset is
seen possibly due to the internal reddening or model uncertainties.  }
\label{fig:z1_cc}
\end{figure}

From a glance at this diagram, it is obvious that the fraction of blue
galaxies is large.  We estimate the blue fraction from this diagram using a
magnitude cut at $z'$=23.2 ($\sim$$M_{V,\rm rest}^*$+1) and a blue/red
separation at $R-z'$=1.3 ($\Delta$($B$$-$$V$)$_{\rm rest}$$\sim$0.2), which
mimic the original Butcher \& Omeler (1984) criteria. We find the blue
galaxy fraction to be as high as $\sim$50\%, which is twice as large as the
one for rich clusters at $z\sim0.5$ suggested by Butcher \& Oemler.
These blue galaxies are probably still vigorously forming stars at $z\sim1$,
and in fact if we plot a field-corrected colour--colour diagram
(Fig.~\ref{fig:z1_cc}), we find that they follow the expected star
formation sequence at $z\sim1$.  A remarkable fact is the complete absence
of galaxies in the regions where we expect foreground/background
contaminants to lie, which strengthens the reliability of our statistical
field correction.  For comparison, the colour--magnitude and colour--colour
diagrams for galaxies in the control fields are shown in
Figs.~\ref{fig:field_cm} and \ref{fig:field_cc}.

We also notice that the scatter in Fig.~\ref{fig:z1_cm} around the red
colour--magnitude sequence is large.  However, we attribute this to the
fact that the galaxies span a range of redshifts ($\Delta$$z$$\sim$0.2).
In fact, the individual colour--magnitude diagrams of the target fields
c1--c5 show tighter red sequences (cf.\ Fig.~\ref{fig:example}).

Fig.~\ref{fig:z1_cm}) also shows the highlight of our results, namely the
colour distribution of galaxies as a function of luminosity.  Specifically,
there are two critical magnitudes at $z'$=21.7 ($M^*-0.5$) and $z'$=24.2
($M^*$+2) which separate this diagram into three characteristic magnitude
ranges: the brightest end ($z'<21.7$) is dominated by red galaxies
($R-z'>1.5$), while the faintest end ($z'>24.2$) is dominated by blue
galaxies ($R-z'<1.5$), and between these is the transition region where
both red and blue galaxies co-exist.  Note that the lower critical
magnitude of $z'$=24.2 is at least one magnitude brighter than the limit of
the data (shown by the vertical dot-dashed line, corresponding to
5--8$\sigma$ detections), assuring $>$10$\sigma$ detections.  The $R$-band
images are also deep enough to detect the red galaxies (see the slanted
dot-dashed line). The deficit of the red faint galaxies we discover is
therefore not caused by incompleteness.

These critical magnitudes correspond to 8$\times$10$^{10}$M$_{\odot}$ and
10$^{10}$M$_{\odot}$, respectively (we describe how we convert from
luminosity to mass in \S4.3).  Therefore, the most luminous (massive)
galaxies are always old while the faintest (least massive) galaxies are
young or still forming significant amount of stars, with the transition
occurring in between.

Bell et al.\ (2004) showed from COMBO-17 data (Wolf et al.\ 2003) that
field galaxies display clear bimodal colour distribution at all redshifts
$z\lsim1$. At $z\sim1$, Bell et al.\ reached only $M^*$, which is 2
magnitudes too bright to see the deficit of red faint galaxies. In this
paper, our deep imaging data have allowed us to find that the bimodal
colour distribution has a strong luminosity dependence at $z\sim1$. The
bimodality is now better described by the existence of two characteristic
populations in the colour--magnitude diagram, namely, `red + bright' and
`blue + faint' populations.  We should note here that Baldry et al.\ (2004)
have recently presented a similar result for the local Universe, based on
SDSS data.

Our interesting results on the colour--magnitude distribution
of galaxies at high redshift ($z\sim1$) will be further discussed
in the following sections based on the colour-dependent luminosity
and stellar mass functions.

\begin{figure}
\begin{center}
  \leavevmode
  \epsfxsize 1.0\hsize
\end{center}
\caption{
A colour--magnitude diagram for the combined control fields with the same
area size as Fig.~\ref{fig:z1_cm} for comparison.
Nearly 5000 galaxies are plotted.
}
\label{fig:field_cm}
\end{figure}

\begin{figure}
\begin{center}
  \leavevmode
  \epsfxsize 1.0\hsize
\end{center}
\caption{
The same as Fig.~\ref{fig:field_cm}, but a colour--colour diagram for the
combined control fields.
}
\label{fig:field_cc}
\end{figure}

\subsection{Luminosity functions}

\begin{figure}
\begin{center}
  \leavevmode
  \epsfxsize 1.0\hsize
  \epsffile{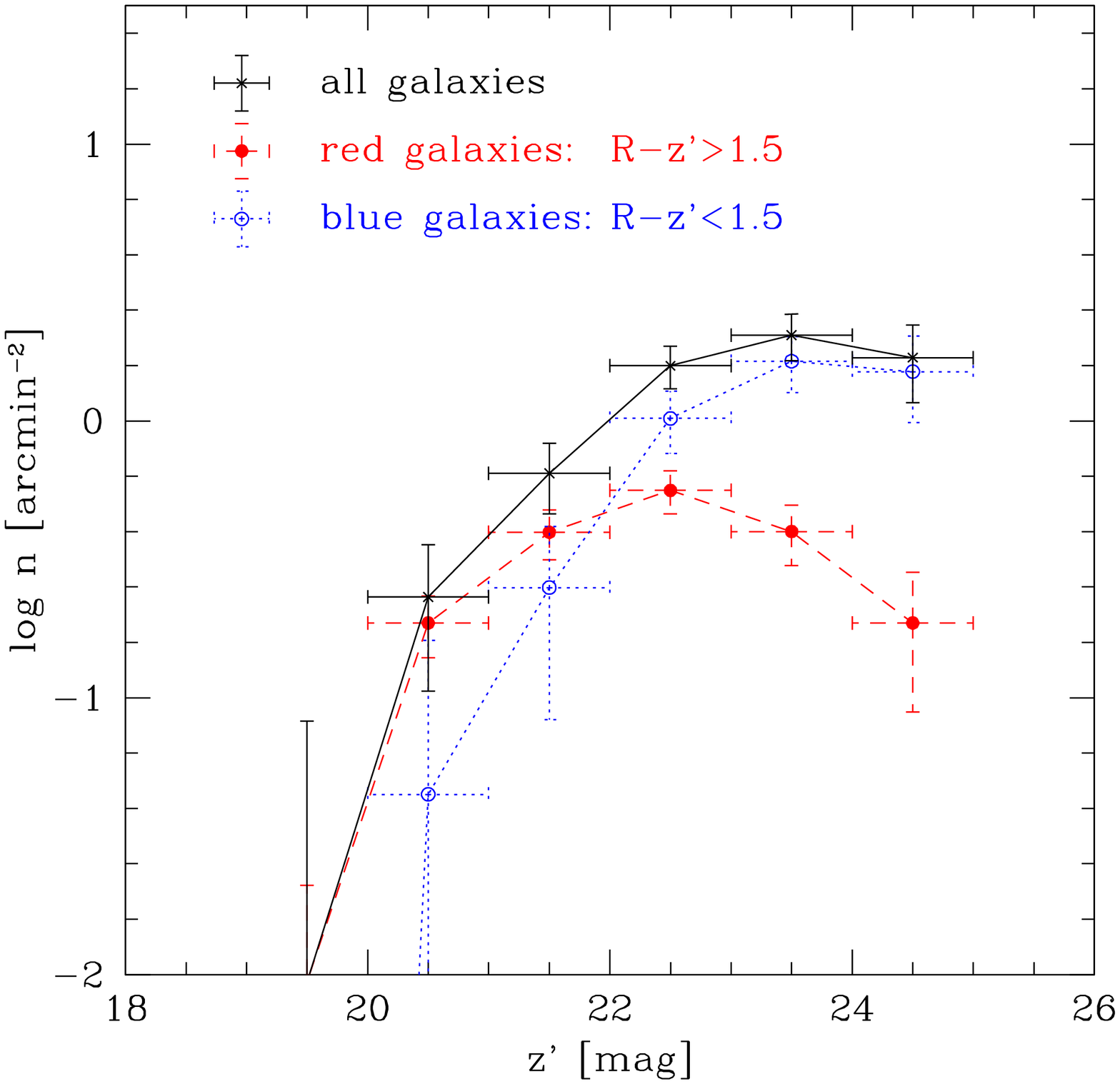}
\end{center}
\caption{
Field-corrected $z'$-band luminosity functions for the $z\sim1$ galaxies.
The solid curve shows the total luminosity function, while the dashed and
dotted curves show the luminosity functions for red and blue galaxies,
respectively, separated at $R-z'=1.5$.  The error bars shown here are
purely Poissonian.  }
\label{fig:lf_z1}
\end{figure}

The field-corrected luminosity functions are shown in Fig.~\ref{fig:lf_z1}.
The overall shape of the luminosity function for all galaxies is a normal
Schechter type although the flattening of the faint end is seen below
$z'\sim23$ ($\sim$$M^*$+1).  It is obvious, however, if the luminosity
function is separated into red and blue galaxies at $R-z'=1.5$ as shown by
the red and blue curves, respectively, they are quite different: the red
one has a bell-like (Gaussian) shape and dominates at brighter magnitudes,
while the blue one has a much steeper slope and dominates the faint number
counts.  Moreover, the deficit of red galaxies at the faint end and the
deficit of blue galaxies at the bright end that we discussed in the
previous subsection are both identified here too.  Note that the faintest
magnitude plotted here is $z'=25$ mag which corresponds to 6--10$\sigma$ in
detection to assure good completeness.  The raw galaxy number counts also
show that the completeness is high at this magnitude (Yamada et al.\ 2003),
which can also be judged from Fig.~\ref{fig:field_cm}.  These
colour-dependent luminosity functions look very similar to the local
luminosity functions derived from the SDSS data (Blanton et al.\ 2001).

Kajisawa et al.\ (2000) claimed a deficit of faint galaxies below
$M^*$+1.5, {\it including both red and blue galaxies}, in the core of the
3C~324 cluster at $z=1.21$ from their deep $K'$-band data.  However, we do
not see such a trend, either in our data (as shown by the solid line in
Fig.~\ref{fig:lf_z1} all the way down to $M^*$+3), nor in the data of
Kodama \& Bower (2003), who presented the $K_s$-band luminosity functions
of $z\sim1$ clusters.  Nakata et al.\ (2001) failed to confirm a deficit in
the cluster member candidates selected from photometric redshifts in their
re-analysis of the 3C~324 cluster. We speculate that the reason why
Kajisawa et al.\ (2000) saw such a deficit of less massive galaxies is
because of their limited field coverage (cluster-centric distance
$r_c$$<$40$''$ or 0.33~Mpc) and hence poor statistics and strong luminosity
segregation.

\subsection{Stellar mass functions}

\begin{figure}
\begin{center}
  \leavevmode
  \epsfxsize 1.0\hsize
  \epsffile{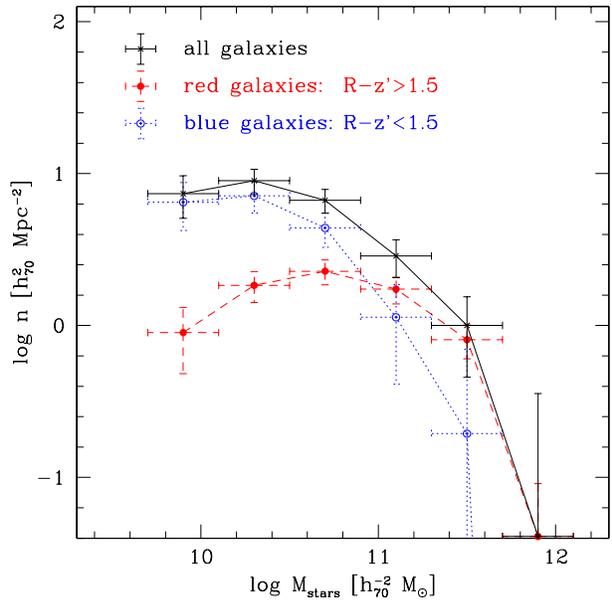}
\end{center}
\caption{
Field-corrected stellar mass functions for the $z\sim1$ galaxies.  The
solid curve shows the total mass function, while dashed and dotted curves
show the mass functions for red and blue galaxies, respectively, separated
at $R-z'=1.5$.  The error bars shown here are purely Poissonian, and the
errors due to field-to-field variation are shown later in
Fig.~\ref{fig:mf_z1_var}.  }
\label{fig:mf_z1}
\end{figure}

Since the $z'$-band still samples rest-frame optical light longer than the
4000\AA\ break for our $z\sim1$ galaxies, we can obtain a rough estimate of
the stellar mass from the $z'$-band luminosity (e.g., Dickinson et al.\
2003).  To do this, however, we need an estimate of the mass-to-light (M/L)
ratio for individual galaxies, which will depend on their star formation
histories.  In fact the M/L ratio in the observed $z'$-band at $z\sim1$
differs by a factor of $\sim$7--8 between passively evolving galaxies
formed at high redshift ($z_f$=5) and those galaxies which have formed
stars constantly over a long period.  Here we use the $R-z'$ colour to
estimate the M/L ratio in the same way as Kodama \& Bower (2003).  In
short, we take the relation between $R-z'$ and M/L ratio at $z=1$ for
models representing a sequence of galaxies with different ratios of bulge
to total mass (Kodama, Bell \& Bower 1999), where a disk component is added
to the underlying passively evolving bulge component.  We then use
Kennicutt's (1983) initial mass function (IMF) to scale the stellar mass.
The resultant field-corrected stellar mass functions of galaxies at
$z\sim1$ are shown in Fig.~\ref{fig:mf_z1}.  Again, the deficit of blue
massive galaxies and the deficit of red less massive galaxies are clearly
seen.

\begin{figure}
\begin{center}
  \leavevmode
  \epsfxsize 1.0\hsize
  \epsffile{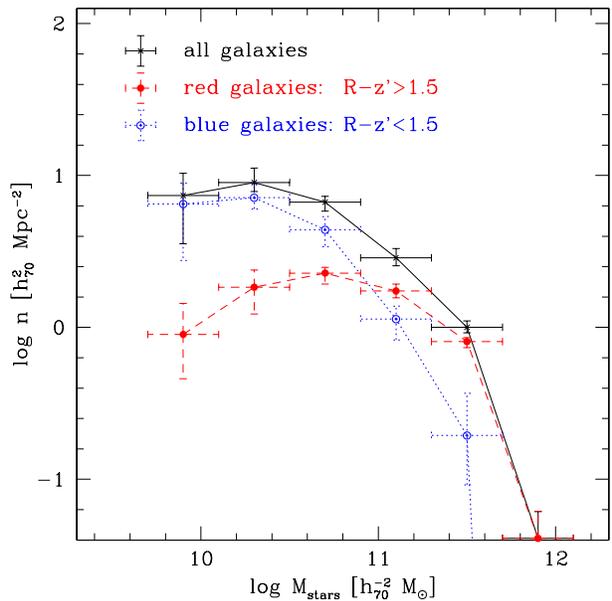}
\end{center}
\caption{
Effect of field-to-field variation in the derived stellar mass functions,
estimated by taking three different sets of control fields.  The upper and
the lower ends of the vertical error bars show the maximum and the minimum
cases while the central dots indicates the average of the three cases.  The
meaning of the curves is the same as in Fig.~\ref{fig:mf_z1}.  }
\label{fig:mf_z1_var}
\end{figure}

The error bars already include the Poisson noise inherent in the limited
number statistics and the statistical field subtraction.  However, the
field subtraction process may also suffer from cosmic variance where galaxy
distribution are known to be non-uniform both locally and at high redshifts
(e.g., Peacock et al.\ 2001; Kodama et al.\ 2001; Shimasaku et al.\ 2003).
Therefore it is essential to cover as wide a field as possible in order to
average out the field-to-field variations.  The wide field coverage of our
data (i.e., 1.2 deg$^2$, corresponding to the co-moving volume of
2.1$\times$10$^6$ Mpc$^3$ within $0.9\le z\le 1.1$) helps us in this
respect.  This also allows us to estimate the effect of field-to-field
variations on $\sim$10~Mpc scales in the derived stellar mass function
within our data.  To achieve this, we take three different and independent
sets of control fields, each consisting of five circled areas with $3'$
radii, giving 141 arcmin$^2$ each. This is the same area as the combined
high density regions from which our $z\sim1$ galaxies are extracted (see
also \S3.2).  Changing the control field populations among these three
independent samples when we perform the field subtraction therefore gives
us an estimate of the internal field-to-field variation on our results.  As
shown in Fig.~\ref{fig:mf_z1_var}, the effect is small and typically
comparable to the pure Poisson errors shown in Fig.~\ref{fig:mf_z1}.  The
deficits of massive blue galaxies and smaller red galaxies are still
significant, and we conclude that field-to-field variation among our
control fields is not a critical issue in our analyses.

\begin{figure}
\begin{center}
  \leavevmode
  \epsfxsize 1.0\hsize
  \epsffile{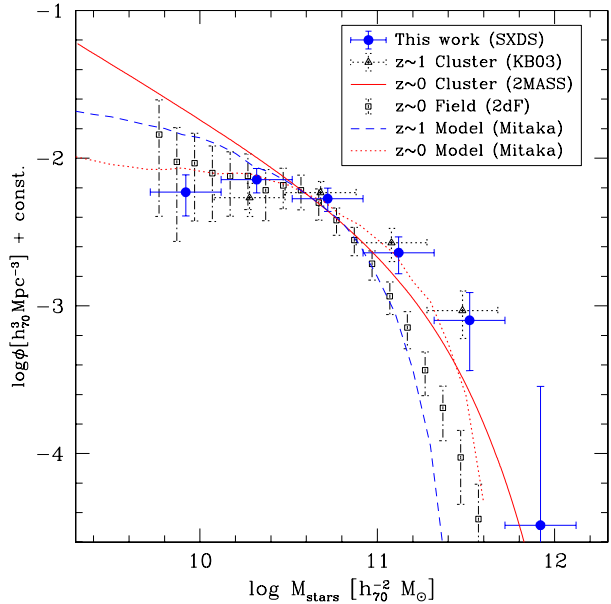}
\end{center}
\caption{
Comparison of our stellar mass functions (blue filled circles with error
bars) with other data and models.  The $z\sim1$ cluster data (open
triangles) are reproduced from Kodama \& Bower (2003) and the 2MASS local
cluster data (solid curve) are taken from Balogh et al.\ (2001).  The 2dF
field data (open squares) are taken from Cole et al.\ (2000).  The model
predictions by the Mitaka semi-analytic model (Nagashima et al.\ 2002) are
shown for $z=0$ (dotted curve) and $z=1$ (dashed curve).  These
predictions are made for cluster galaxies taken from dark haloes whose
circular velocities are larger than 1000~km/s.  The observed data and the
models are all normalised at 5$\times$10$^{10}$M$_{\odot}$ to have the same
amplitude as the 2dF mass function.  }
\label{fig:mf_comp}
\end{figure}

It is also interesting to examine the evolution of the stellar mass
functions with redshift since this provides a critical test for the
CDM-based bottom-up scenario for galaxy formation and evolution (e.g.,
Kauffmann \& Charlot 1998; Baugh et al.\ 2002).  In Fig.~\ref{fig:mf_comp},
our total stellar mass function (blue filled circles) is compared to
previous measurements of the stellar mass function at $z\sim1$ and
$z\sim0$.  All the curves and data points have been normalised at
5$\times$10$^{10}$M$_{\odot}$ to have the same amplitude as the 2dF mass
function.  The agreement with the stellar mass function of $z\sim1$ cluster
galaxies based on near-infrared imaging (Kodama \& Bower 2003) is
remarkable, which suggests that we are indeed looking at high density
regions at $z\sim1$ in this study.  If we compare the observed stellar mass
functions at $z\sim1$ to their lower redshift counterpart (e.g., 2MASS
clusters; Balogh et al.\ 2001), we see the lack of evolution from $z\sim1$
to the present-day, as discussed in detail by Kodama \& Bower (2003)
(see also Kodama et al.\ 2004 for extension to even higher redshift
of $z\sim1.5$).
There are very massive galaxies ($>$10$^{11}$M$_{\odot}$) already in place
in high density regions at $z\sim1$, which must therefore have been
assembled even earlier.  The rapid mass assembly of these large galaxies is
contrasted to the semianalytic model predictions of Nagashima et al.\
(2002) which shows a much slower mass assembly for cluster galaxies between
$z=1$ and $z=0$ (blue dashed and red dotted lines in
Fig.~\ref{fig:mf_comp}, respectively).

A deviation can be seen at the faint end between the local clusters (2MASS)
and the $z\sim1$ high density regions (this work and Kodama \& Bower 2003)
where the slope is flatter in the high-$z$ systems.  However, the faint end
slope of the 2MASS clusters is only poorly determined (Balogh et al.\
2001), and we do not discuss this issue further.

\section{Discussion}

\subsection{Down-sizing in galaxy formation}

The most important result shown in this paper is that galaxy formation
processes, both mass assembly and star formation, take place rapidly and
are completed early in massive systems, while in the less massive systems,
these processes (at least star formation) are slower.  This `down-sizing' in
galaxy formation was first noted by Cowie et al.\ (1996) who observed that
the maximum luminosity of galaxies hosting star formation decreases with
decreasing redshift at $z\gsim1$, although their data (from the {\it Hawaii
Deep Fields\/}) were limited to galaxies brighter than $<M^*$
(see also Pozzetti et al.\ 2003; Kashikawa et al.\ 2003;
Poggianti et al.\ 2004; Tran et al.\ 2004).

Kodama \& Bower (2001) investigated the magnitude dependent evolution of
cluster galaxies from intermediate redshifts $0.2\lsim z\lsim 0.4$ (CNOC
clusters, Yee et al.\ 1996) to the present day (Coma cluster, Terlevich et
al.\ 2001) by letting the distant clusters evolve forward in time by
truncating the star formation in blue galaxies. This simulation showed that
the less massive galaxies seen today had a higher fraction of blue galaxies
in the past compared to the massive galaxies, although the data were
complete only to $\sim$$M^*$+1.  The Butcher--Oemler effect (Butcher \&
Oemler 1984 show a systematic decrease of blue galaxy fraction above
$M_V=-20$ with decreasing cluster redshift) can then be partly explained by
the down-sizing in star forming galaxies.  If these blue galaxies fade and
become red quickly after ceasing star formation, they will enter the faint
end of the red colour--magnitude sequence.  If star formation is truncated
from massive galaxies towards less massive galaxies as time progresses, the
fraction of blue galaxies brighter than a certain magnitude limit should
decrease with time, in line with observations (Bower, Kodama \& Terlevich
1998; Kodama \& Bower 2001).  As the blue galaxies with brighter
luminosities stop their star formation, they will evolve to redder colours
and fainter magnitudes along the iso-mass lines shown by the dashed lines
in Fig.~\ref{fig:z1_cm}, and the faint-end of the red sequence where we see
a deficit of galaxies at $z\sim1$ will be populated with galaxies. This is
important because such a deficit of red galaxies between $M^*$+2 and
$M^*$+3 is not seen locally, at least not in the Coma cluster (Terlevich et
al.\ 2000).

We have discussed a significant gradient in age or star forming activity as
a function of galaxy luminosity or mass.  For clarity, however, we note
that our results are not inconsistent with those of Kodama et al.\ (1998),
who discussed a lack of age gradient for cluster early-type galaxies, since
they studied only morphologically-selected early-type galaxies brighter than
$M^*$+1.  Furthermore, once star formation has terminated, the blue
galaxies would quickly fade and redden and the luminosity weighted age
would hardly change along the red colour--magnitude sequence (Kodama \&
Bower 2001).

What can cause this down-sizing, which is apparently at odds with the
`bottom-up' scenario that is a natural consequence of CDM models.  Biased
galaxy formation (e.g., Cen \& Ostriker 1993), in which massive galaxies
form from the highest peaks in the initial density fluctuation field, may
help to solve this apparent contradiction. The densest fluctuations
collapse earlier and hence galaxy formation, both mass assembly and star
formation, takes place in an accelerated way in these globally biased
regions, while the less massive galaxies seen at $z\sim1$ collapse later on
average since they formed from isolated peaks with lower density. However,
a further puzzle is how to keep the star formation rate low so that it
continues for long enough to still produce blue galaxies at $z\sim1$.  Two
possible mechanisms have so far been proposed to explain this. One
suggestion is that the UV background may penetrate into small systems to
keep them ionized and prevent star formation (Babul \& Rees 1992; Kitayama
et al.\ 2001).  The other idea is that supernova explosions cause a periodic
gas outflow and re-inflow of gas from the shallow potential well over a
long period of time (e.g., Dekel \& Silk 1986).  Yet no theory can explain,
either quantitatively or even qualitatively, the `down-sizing' in galaxy
formation, and this is a big challenge for theorists working in this field.

From the observational side, it would be extremely interesting to directly
measure the evolution in the break mass with time (cf.\ Kodama \& Bower
2001).  Kauffmann et al.\ (2003) and Baldry et al.\ (2004) find a break
mass of $\sim$2$\times$10$^{10}$M$_{\odot}$ in the local Universe (SDSS),
which is roughly the same break mass as we find at $z\sim1$.  However, in
this paper we have discussed only the high density regions, whereas the
local analyses are based on the averaged field properties.  A direct
comparison is therefore not meaningful since galaxy properties are strongly
dependent on environment in the sense that galaxy formation and evolution
processes are accelerated in high density regions (e.g., Dressler et al.\
1980; Kodama et al.\ 2001; Gomez et al.\ 2003; Balogh et al.\ 2004).  In
fact, extending the deep colour--magnitude analyses of galaxies of this
kind along the environmental axis is important since we expect to see a
higher break mass in lower density environments due to slower galaxy
evolution.  It is also appealing to extend this analysis to even higher
redshifts ($z\gg1$) where deep near infrared surveys are essential.  With
the upcoming wide-field near infrared cameras such as MOIRCS
(4$'$$\times$7$'$) on Subaru and WFCAM (51$\times$51 arcmin$^2$ by four
pointings) on UKIRT, we will be able to
probe the era where massive galaxies are in the process of formation.

\section{Conclusions}

In this paper, we have presented a photometric analysis of galaxies in
colour-selected high density regions at $z\sim1$, constructed from the
unique deep ($z'$=25, 6--10$\sigma$) and wide (1.2 deg$^2$) optical
multi-colour imaging data taken as a part of the Subaru/XMM-Newton Deep
Survey (SXDS) project. Our analysis has been based primarily on the
field-corrected colour--magnitude diagram and the colour-dependent stellar
mass functions.

Benefitted by the depth of the survey, we have found a deficit of faint red
galaxies along the colour--magnitude sequence below $M^*$+2 with respect to
the passive evolution, or $\sim$10$^{10}$M$_{\odot}$.  Almost all galaxies
below this luminosity/mass at $z\sim1$ are still undergoing significant
star formation.  The luminous/massive end ($<$$M^*$$-$0.5), however, is
dominated by old red systems with almost no blue galaxies.  The clear
distinction of the `red + bright' and `blue + faint' populations at
$z\sim1$ on the colour--magnitude diagram suggests the existence of
`down-sizing' in galaxy formation, where star formation is switched off
from the massive systems towards the less massive ones as the Universe
ages.  We find that the mass assembly process is also largely complete by
$z\sim1$ in the high density regions.  How to accommodate this down-sizing
phenomena in galaxy formation within the context of the bottom-up scenario
of the CDM Universe is an interesting puzzle.

\section*{Acknowledgments}

We thank Drs Richard Bower, Ian Smail, Eric Bell and Alvio Renzini for
useful discussion.  We are also grateful for the Referee, Dr C. Wolf, for
his constructive comments.  This work is based on data collected at Subaru
Telescope, which is operated by the National Astronomical Observatory of
Japan, and was financially supported in part by a Grant-in-Aid for the
Scientific Research (No.\, 15740126) by the Japanese Ministry of Education,
Culture, Sports, Science and Technology. CS is supported by an Advanced
Fellowship from the Particle Physics and Astronomy Research Council.

\end{document}